\documentclass[journal,twoside,web]{ieeecolor}
\usepackage{tmi}
\usepackage{cite}
\usepackage{amsmath,amssymb,amsfonts}
\usepackage{algorithmic}
\usepackage{graphicx}
\usepackage{textcomp}

\usepackage{graphicx}
\usepackage{multirow}

\usepackage{multirow}
\usepackage{tabu}
\usepackage{svg}
\usepackage{bbm}
\usepackage{diagbox}
\usepackage[english]{babel}
\usepackage{comment}
\usepackage{array}
\usepackage{amstext}
\usepackage{makecell}
\usepackage{caption}
\usepackage{tablefootnote}
\usepackage{babel}
\usepackage{booktabs} 
\usepackage{float}

\usepackage[ruled,vlined]{algorithm2e}
\newcommand{\nosemic}{\renewcommand{\@endalgocfline}{\relax}}
\newcommand{\dosemic}{\renewcommand{\@endalgocfline}{\algocf@endline}}
\usepackage{chngcntr}
\usepackage{placeins}
\usepackage{amsfonts}       
\usepackage{nicefrac} 
\usepackage{amsmath,amssymb} 
\usepackage{amstext}

\newcommand{\ie}{\textit{i.e.}}
\newcommand{\eg}{\textit{e.g.}}

\SetKwInput{KwInput}{Input}                
\SetKwInput{KwOutput}{Output}              
\usepackage[pagebackref=false,breaklinks=true,letterpaper=true,colorlinks,bookmarks=false]{hyperref}

\newlength\mylen

\newcommand{\revised}[1]{{#1}}

\def\BibTeX{{\rm B\kern-.05em{\sc i\kern-.025em b}\kern-.08em
    T\kern-.1667em\lower.7ex\hbox{E}\kern-.125emX}}
\newcommand{\bfsection}[1]{\vspace*{0.1cm}\noindent\textbf{#1.}}

\begin{document}
\title{\emph{RibSeg v2}: A Large-scale Benchmark for Rib Labeling and Anatomical Centerline Extraction}

\author{Liang Jin, Shixuan Gu, Donglai Wei, \revised{Jason Ken Adhinarta}, Kaiming Kuang, \revised{Yongjie Jessica Zhang},\\Hanspeter Pfister, Bingbing Ni, Jiancheng Yang, and Ming Li
    \thanks{This work was supported by Science and Technology Planning Project of Shanghai Science and Technology Commission (22Y11910700), Shanghai Key Lab of Forensic Medicine, Key Lab of Forensic Science, Ministry of Justice, China (Academy of Forensic Science) (KF202113), National Natural Science Foundation of China (61976238), Shanghai 'Rising Stars of Medical Talent' Youth Development Program 'Outstanding Youth Medical Talents' (SHWJRS [2021]-99), Emerging Talent Program (XXRC2213) and Leading Talent Program (LJRC2202) of Huadong hospital, and Excellent Academic Leaders of Shanghai. This work was supported in part by National Science Foundation of China (U20B2072, 61976137) and Shanghai Jiao Tong University Medical Engineering Cross Research (YG2021ZD18) (Corresponding authors: Ming Li, Jiancheng Yang, and Bingbing Ni.)}
    \thanks{Liang Jin and Shixuan Gu contributed equally as co-first authors.}
	\thanks{Liang Jin is with Radiology Department, Huadong Hospital, affiliated to Fudan University, Shanghai, China and with Huashan Hospital, affiliated to Fudan University, Shanghai, China and also with Shanghai Key Lab of Forensic Medicine, Key Lab of Forensic Science, Ministry of Justice, China (Academy of Forensic Science) (jin\_liang@fudan.edu.cn)}
	\thanks{Shixuan Gu is with Carnegie Mellon University, PA, USA, and also with Harvard University, MA, USA (\revised{shixuangu@g.harvard.edu}).} 
	\thanks{Donglai Wei \revised{and Jason Ken Adhinarta} are with Boston College, MA, USA (donglai.wei@bc.edu, \revised{jason.adhinarta@bc.edu)}.}
	\thanks{Kaiming Kuang is with \revised{University of California San Diego, CA, USA}, and also with Dianei Technology, Shanghai, China (\revised{kakuang@ucsd.edu}).}
 	\thanks{\revised{Yongjie Jessica Zhang is with Carnegie Mellon University, PA, USA (jessicaz@andrew.cmu.edu).}}
	\thanks{Hanspeter Pfister is with Harvard University, MA, USA (pfister@seas.harvard.edu).}
	\thanks{Bingbing Ni is with Shanghai Jiao Tong University, and also with Huawei Hisilicon, Shanghai, China (nibingbing@sjtu.edu.cn).}
	\thanks{Jiancheng Yang is with Shanghai Jiao Tong University, and also with EPFL, Lausanne, Switzerland (jekyll4168@sjtu.edu.cn).}
	\thanks{Ming Li is with Radiology Department, Huadong Hospital, affiliated to Fudan University, Shanghai, China, and also with Institute of Functional and Molecular Medical Imaging, Shanghai, China (e-mail: minli77@163.com).}
}

\maketitle

\begin{abstract}
Automatic rib labeling and anatomical centerline extraction are common prerequisites for various clinical applications. Prior studies either use in-house datasets that are inaccessible to communities, or focus on rib segmentation that neglects the clinical significance of rib labeling. To address these issues, we extend our prior dataset (\emph{RibSeg}) on the binary rib segmentation task to a comprehensive benchmark, named \emph{RibSeg v2}, with 660 CT scans (15,466 individual ribs in total) and annotations manually inspected by experts for rib labeling and anatomical centerline extraction. 
Based on the \emph{RibSeg v2}, we develop a pipeline including deep learning-based methods for rib labeling, and a skeletonization-based method for centerline extraction. To improve computational efficiency, we propose a sparse point cloud representation of CT scans and compare it with standard dense voxel grids. Moreover, we design and analyze evaluation metrics to address the key challenges of each task. Our dataset, code, and model are available online to facilitate open research at \hyperlink{https://github.com/M3DV/RibSeg}{https://github.com/M3DV/RibSeg}.

\end{abstract}


\section{Introduction}
\label{sec:introduction}

\IEEEPARstart{R}{ib} labeling and anatomical centerline extraction are of significant clinical value to facilitate various clinical applications. For example, it is critical for detecting rib fractures, which can identify chest trauma severity that accounts for $10\%\sim15\%$ of all traumatic injuries~\cite{sirmali2003comprehensive}. Besides, the structure and morphology of rib bones are stable references for multiple analysis and quantification tasks such as lung volume estimation~\cite{Mansoor2014AGA,Xu2014EfficientRS}, bone abnormality quantification~\cite{Fokin2018QuantificationOR} and pediatric spinal deformities~\cite{Tajdari2021AIS,Tajdari2022PSD}. Based on rib anatomical centerlines, internal coordinate systems can localize organs for surgery planning and postoperative evaluation~\cite{Wang2005ART}, as well as registering pathologies such as lung nodules~\cite{Shen2003ATC}. Moreover, automatic rib labeling and centerline extraction is the key to developing visualization tools of unfolded rib cages~\cite{ringl2015ribs,Bier2015EnhancedRT, Abe2014HighspeedPC}, significantly reducing the burden of rib interpretation for clinicians. 

\begin{figure}[tb]
    \centering
	\includegraphics[width=\linewidth]{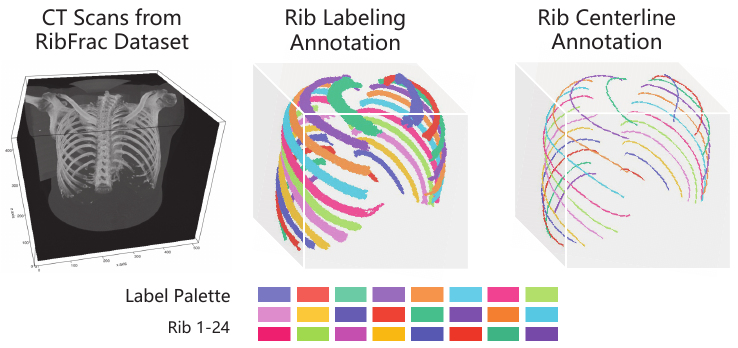}
	\caption{\textbf{\emph{RibSeg v2} Dataset}. \emph{RibSeg v2} extends the annotations of 660 CT scans from the existing \emph{RibFrac} dataset~\cite{jin2020deep}, containing labeled rib segmentation and anatomical centerlines. The color palette indicates the label assigned to each rib.} \label{fig:RibSeg_overview}
\end{figure}

Despite the high contrast of ribs, rib labeling and centerline extraction is challenging. Ribs in human bodies are typically elongated and oblique across numerous CT sections; In other words, a large number of CT slices must be evaluated sequentially by radiologists. Ribs are anatomically close to the scapula and clavicle, and some ribs might be connected by the metal implant, which is hard to label. Furthermore, extraction of the rib centerline is subject to image noise, artifacts, and the quality of rib labels.

Previous studies on this topic only focus on rib segmentation~\cite{staal2007automatic} and trivialize rib labeling as a counting process~\cite{Wu2021DevelopmentAE,lenga2018deep}, which underestimates the challenges of false merging and neglects that anatomical labeling of ribs is clinically more desirable. Tracing-based rib segmentation and centerline extraction methods are highly sensitive to initially detected seed points and vulnerable to local ambiguities\cite{shen2004tracing,klinder2007automated}. Although the deep learning-based method is robust as it learns hierarchical visual features from raw voxels\cite{wu2012learning}, it does not consider the sparsity and elongated geometry of ribs. Moreover, there is no public dataset on this topic, making it difficult to benchmark existing methods and develop downstream applications such as image-based modeling and simulations~\cite{Zhang2013,Zhang2016Book}.






To tackle these problems, we first develop a benchmark for rib labeling and anatomical centerline extraction, named \emph{RibSeg v2}, including manually inspected annotations of 660 chest-abdomen CT scans (15,466 individual ribs) from \emph{RibFrac} dataset\cite{jin2020deep}. In addition, we formulate rib labeling as the task of segmenting ribs from CT scans and labeling the binary segmentation into 24 instances, and benchmark the \emph{RibSeg} with a pipeline including a deep learning-based method for rib labeling and skeletonization-based~\cite{teasar} methods for rib anatomical centerline extraction. We further compared the data representations of CT scans as dense voxel grids and sparse point clouds, respectively, and proposed various metrics for each task to perform comprehensive evaluations.

This study is extended from our \emph{RibSeg} (v1) previously presented at MICCAI~\cite{yang2021ribseg}, where we introduced a binary rib segmentation benchmark with 490 CT scans from the RibFrac dataset~\cite{jin2020deep}. 
In this study, we extend \emph{RibSeg} to a comprehensive benchmark for \emph{rib labeling} and \emph{anatomical centerline extraction} by adding 
1) the 170 remaining cases with binary rib segmentation, which are hard to be segmented by the semi-auto method~\cite{yang2021ribseg}, 
2) rib labels (1-24) except for the \revised{6} unqualified cases, and 
3) annotations of rib anatomical centerlines except for the \revised{6} unqualified cases. All annotations are manually checked, and CT scans that are hard to annotate are categorized into challenging cases by 2 junior radiologists based on visual assessment. Besides, we extended the previous method for binary rib segmentation to rib labeling and anatomical centerline extraction. Finally, by detailed quantitative and qualitative analysis of the challenging cases, we explored the key challenges of each task, which are valuable to facilitate future studies on this topic.


\section{Related Works}
\subsection{Automatic Rib Analysis}
\bfsection{Rib segmentation and labeling} 
A few studies have addressed rib segmentation and labeling\cite{shen2004tracing,klinder2007automated} before the era of deep learning, where rib tracing with initial seed point detection is the key method. Supervised deep learning-based segmentation~\cite{Wu2021DevelopmentAE} from CT volumes is robust, as it adopts 3D-UNet\cite{cciccek20163d} to learn hierarchical visual features from raw voxels. MDU-Net\cite{Wang2020MDUNetAC} is proposed to segment clavicles and ribs from CT scans, which combines multiscale feature fusion with the dense connection\cite{Huang2017DenselyCC}.

\bfsection{Rib anatomical centerline extraction} 
A few non-learning studies work on rib centerline extraction by modeling the ribs as elongated tubular structures and conducting rib voxel detection by structure tensor analysis\cite{Aylward2002tubular,staal2007automatic}. Rib tracing-based method is also introduced to centerline extraction\cite{ramakrishnan2011automatic}. There are also deep learning-based studies focusing on rib centerline extraction instead of full rib segmentation, \eg, rib centerlines are extracted by applying morphological methods such as deformable template matching\cite{wu2012learning} and rib tracing method\cite{lenga2018deep} to the rib cages detected by deep learning method.

\subsection{Deep Learning Models for 3D CT Volumes}

\revised{Most studies model CT scans as 3D volumes, and work on dense voxel grids, which is computationally expensive. In this study, we represented CT scans as dense voxel grids and sparse point clouds, respectively, for the method comparison.}

\bfsection{Voxel grids} 
3D-UNet\cite{cciccek20163d} is first introduced to work on sparsely annotated volumetric data. VoxSegNet\cite{Wang2020VoxSegNetVC} is further proposed as an effective volumetric method for 3D shape part segmentation, which extracts discriminative features encoding detailed information under limited resolution. PVCNN\cite{Liu2019PointVoxelCF} and PointGrid\cite{Le2018PointGridAD} integrate representations of points and voxels to enhance feature extraction and model efficiency. 

\bfsection{Point clouds} 
Deep learning for point cloud analysis\cite{guo2020deep} is pioneered by PointNet\cite{qi2017pointnet} and DeepSet\cite{zaheer2017deep}. Later studies also introduce sophisticated feature aggregation based on spatial graphs\cite{qi2017pointnet++,liu2019dynamic} or attention\cite{yang2019modeling}. In medical imaging scenarios, point cloud matching has been applied to 3D volumes\cite{Abe2014HighspeedPC} since 2014. The transformer mechanism\cite{Vaswani2017AttentionIA} is further introduced for medical point cloud analysis\cite{Yu20213DMP}, and point cloud-based methods are adapted to various medical applications such as nodule detection\cite{Drokin2020DeepLO} and vessel reconstruction\cite{Banerjee2020PointCloudMF}.

\subsection{Semantic Segmentation-guided Methods}
\revised{The extreme foreground-background imbalance and
data sparsity are common challenges of part segmentation tasks in (bio)medical domains. The semantic segmentation-guided method is a common solution, which is also widely used in fine-grain classification tasks such as pedestrian detection\cite{Liu2018FasterRF,Yu2019PedestrianSB} where semantic segmentation is first performed to obtain complementary higher-level semantic features.} In this study, considering the sparsity of ribs in CT volumes, we first perform foreground-background segmentation to roughly segment the ribs and label them by multi-class segmentation with a second model. A similar pipeline is also used in the medical scenario, such as intracranial aneurysm segmentation\cite{Yang2020ATS}, where vessel segments with aneurysms are detected from the whole CT scan, and segmented by a second model. This pipeline essentially addresses the sparsity issue and eases follow-up tasks.

\subsection{Skeletonization Methods} 
\revised{In this study, rib anatomical centerline is extracted from well-segmented rib labels, which can be essentially formulated as a skeletonization task for elongated objects.}

\bfsection{Learning-based skeletonization}
Most studies of learning-based skeletonization work on 2D images, \eg, DeepFlux predicts a two-dimensional vector field to map scene points to extract the skeleton\cite{Wang2019DeepFluxFS}, and the skeleton can also be extracted by integrating image and segmentation to obtain complementary information\cite{Liu2017FusingIA}. For 3D skeleton extraction, the previous study utilizes normalized gradient vector flow on volume data\cite{Yoon20093DSE}, and most studies of 3D skeleton focus on human recognition and re-identification\cite{Rao2021ASG}, \eg, PointSkelCNN is proposed to extract 3D human skeleton from point clouds\cite{Qin2020PointSkelCNNDL}.

\bfsection{Voxel-based TEASAR method} 
The \emph{Tree-structure Extraction Algorithm for Accurate and Robust Skeletons} (TEASAR)\cite{teasar, Zhao2018neutu} is originally proposed to skeletonize binary discretized 3D occupancy maps of tree-like structures, such as neurons\cite{Fornito2013connectome,Pawar2020}. The pipeline of the original TEASAR is summarized as follows: 1) first locate a root point on the rib volume, 2) and then serially trace the shortest path via a penalty field\cite{Bitter2001penalizedis} to the most distant unvisited point. 3) After each passing, a circumscribing cube is applied to expand around the vertices in the path, marking the visited regions. 4) Repeat the process above until the whole volume is traversed.

\revised{
\bfsection{Point-based L1-medial skeletonization}
L1-medial skeletonization~\cite{Huang2013L1medialSO} is well-known as a state-of-the-art method to extract curve skeleton for point clouds. It used L1-median as a robust global center of the point cloud, and by adapting L1-medians locally to a point set represeneting a 3D shape, the resultant 1D structure can serve as a localized center of the shape, \ie, the centerline. 
}


\section{\emph{RibSeg V2} Dataset}
\subsection{Dataset Overview}
Most prior studies on rib segmentation or rib centerline extraction use small in-house datasets\cite{Wang2020MDUNetAC}, which makes it inconvenient to conduct comparative studies and develop new methods. To address this issue, we developed the \emph{RibSeg} (v2) dataset containing 660 cases, with 15,466 ribs in total. Considering the clinical practicality, we further categorize the cases that are hard to annotate as challenging cases. Fig.~\ref{fig:RibSeg_overview} gives an overview of the \emph{RibSeg v2} dataset. 

\bfsection{Data source}
\emph{RibSeg v2} Dataset uses the public computed tomography (CT) scans from the \emph{RibFrac} dataset~\cite{jin2020deep}, an open dataset with 660 chest-abdomen CT scans for rib fracture segmentation, detection, and classification. The CT scans are saved in NIFTI (.nii) format with volume sizes of $N \times 512 \times 512$, where $512 \times 512$ is the size of CT slices, and $N$ is the number of CT slices (typically $300\sim500$). Most cases are confirmed with complete rib cages (24 ribs) and manually annotated with at least one rib fracture by senior radiologists. 

\begin{table}[tb]
	\caption{\textbf{Data Division and Stats of \emph{RibSeg v2} Dataset}. The table includes the number of total cases, individual ribs, cases with the incomplete rib cage, and unqualified cases for each subset. The unqualified cases refer to the cases that 1) miss annotations of labels or centerlines, and 2) have flaws in rib label annotations. The file names and details of these abnormal cases are categorized into a dataset description file, which will be made available together with \emph{RibSeg v2} dataset.} \label{tab:data-overview}
	\centering
	\begin{tabular*}{\hsize}{@{}@{\extracolsep{\fill}}lcccc@{}}
		\toprule
		\multirow{2}{*}{Subset} & \multirow{2}{*}{CT Scans} & \multirow{2}{*}{Individual Ribs} & Incomplete & Unqualified \\
		& & & Rib Cages & Cases \\
		\midrule
		Training & 420 & 9,961 & 28 & \revised{0} \\
		Development & 80 & 1,780 & 13 & \revised{6} \\
		Test & 160 & 3,725 & 32 & \revised{0} \\
		\bottomrule
	\end{tabular*}
\end{table}

\bfsection{Dataset division and statistics}
The data split of the \emph{RibSeg v2} dataset is summarized in Tab.~\ref{tab:data-overview}: training set (420 cases), development set (80 cases), and test set (160 cases). The division of \emph{RibSeg v2} training, development, and test sets are from those of the \emph{RibFrac} dataset respectively, facilitating the development of downstream applications such as rib fracture detection. In Tab.~\ref{tab:data-overview}, we also report the number of cases with incomplete rib cages and unqualified cases. Specifically, the cases with incomplete cages only cover the upper chest-abdomen region, while the unqualified cases refer to the cases whose annotations are missed or contain potential flaws, including 4/4/3 (training/development/test) cases that miss annotations of labels or centerline due to the CT scans quality degradation, and 23/5/8 cases with label crossing in annotations. The file names and details of all the abnormal cases are categorized into a dataset description file, which will be made available together with the \emph{RibSeg v2} dataset.


\begin{figure*}[tb]
    \centering
	\includegraphics[width=\linewidth]{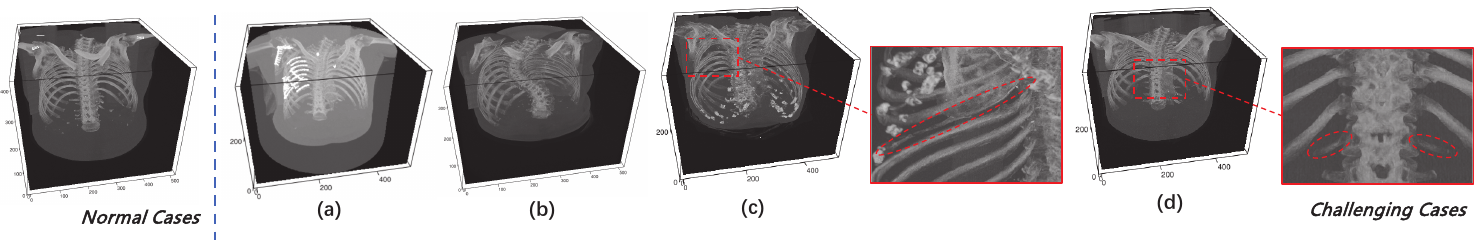}
	\caption{\textbf{Challenging Cases}. 76 cases in \emph{RibSeg v2} are categorized as challenging cases. a) The case suffers HU deviation, and the left 3rd$\sim$9th ribs are connected by metal plates. b) The case contains serious scoliosis. c) The case contains serious fractures in the sternum and the left 5th rib. d) The floating ribs are abnormally short (the 12th pair of ribs).} \label{fig:challenging_cases}
\end{figure*}

\subsection{Data Annotation}
Annotating rib labels and anatomical centerlines from CT scans is labor-intensive due to the elongated and oblique shape of ribs. To ease the workload and facilitate the annotation, we develop a morphological pipeline to obtain rib label segmentation\cite{yang2021ribseg}. Based on the high-quality labels, we apply skeletonization methods to extract anatomical centerlines. For abnormal cases where the pipeline fails, we first \revised{manually annotate the rib centerlines, and then apply a series of morphological operations to obtain rib label segmentation based on the upsampled centerlines.} Each step contains manual checking and refinement, and final annotations are confirmed by \revised{2 junior radiologists and 1 senior radiologist with a human-in-the-loop procedure}.

\bfsection{Rib labeling}
Rib labeling contains segmenting ribs from CT scans and labeling the segmentation. \revised{To generate binary segmentation, we initially adopted the semi-automatic pipeline\cite{yang2021ribseg} introduced in \emph{RibSeg} (v1) and successfully produced 519 cases, including 480 cases from \emph{RibSeg} (v1).} We describe the primary steps as follows: For each volume, we first filter out non-target voxels by thresholding at 200 HU\cite{Lev200217C} and then separate the ribs from the vertebra through morphological methods (\eg, dilation, and erosion). For cases that remain parts of the clavicle and scapula, we manually locate and remove them according to the coordinates of their connected components\cite{Rosenfeld1966dip,Wu2005ccl}. The resultant rib segmentation is then labeled from top to bottom and left to right. 

\revised{For the rest 127 cases where it fails, we turn to a centerline-based pipeline. 1) We first have \revised{1} senior radiologist manually annotate the centerlines. Specifically, the radiologist would annotate 10 points within each rib through meticulous inspection of the CT slices, and the points are further interpolated and smoothened into a 3D curve of 500 points as the centerline. 2) For each rib, we dilate its centerline as a mask and then take the overlap of its mask and CT volume as the rough segmentation. 3) For the resultant rough segmentation, we filter the noise voxels by keeping the largest connected components. In cases where severe fracture breaks the rib into multiple volumes, we adjust the number of the connected components to be kept based on the visual assessment of the resultant segmentation. 4) To ensure the completeness of annotation, we repeatedly dilate the rib segmentation, take the overlap of it and CT volume, and filter the noise voxels. 

This pipeline generates high-quality rib labels for 121 cases, and all results are manually checked and refined in a human-in-the-loop procedure to ensure high quality. The rest 6 cases, however, are partly scanned and only cover the middle part of the ribcage, which makes it impossible to label the ribs. Hence, we only provide the annotations of binary segmentation and unlabeled centerline for these cases, and denote them as unqualified in Tab.~\ref{tab:data-overview}, and categorized into a dataset description file, which will be made available together with the \emph{RibSeg v2} dataset. Note that these cases are primarily used for lung inspection, and do not contain rib fractures. We previously incorporated these cases into the evaluation and test sets of the \emph{RibFrac} dataset in order to assess the model's robustness.}

\bfsection{Rib anatomical centerline extraction}
\revised{Annotating a 3D centerline by manually inspecting 2D slices is a naturally challenging task even for well-trained radiologists. This is because it is nearly impossible to locate endpoints that lie precisely in the center of the rib based on 2D perception, as the thickness of the rib is difficult to assess. Even for the 127 cases of centerlines that are manually annotated by radiologists, they might deviate from the center of the ribs. Hence, for all 660 cases, based on the manually confirmed rib segmentation, we extract the rib anatomical centerlines by implementing 1) a variant of the voxel-based TEASAR method, and 2) a point-based L1-medial skeletonization method. For each rib, we compare the results of these 2 skeletonization methods (and manual annotation if contains) based on both visual assessment and numerical analysis. 

Specifically, the centerlines generated by skeletonization methods could be tortuous since the rib components are hollow inside, and the sizes of the point set composing centerlines are different. Hence, we post-process the centerlines by smoothening and upsampling them to 500 points. Then we first evaluated all the results based on visual assessment. When two skeletonization-based results (and manual annotation) are visually similar, we calculated the Chamfer distance~\ref{eq:lscd} between the centerlines and rib segmentation as a reference for the deviation of the centerline from the actual centerline of the rib. Then we selected the one with the lowest distance as the finalized centerline. And for the ribs where skeletonization-based methods failed, we used the smoothened manual annotations as the finalized centerline.}

\bfsection{Manual proofreading}
The abnormal cases, along with the pursuit of high annotation quality, incentivize us to perform laborious checking and refinement after the annotation stages. For instance, in quite a few cases, the floating ribs are too short or sparse that the segmentation vanished after the morphological procedure. Hence, we manually check and refine the annotation case by case. To recover and annotate missed ribs, we turn back to manually ensure the segmentation completeness by modifying the corresponding connected components voxel by voxel. To ensure high quality, all final segmentations and centerlines are manually checked, refined, and confirmed by \revised{2 junior radiologists and 1 senior radiologist} based on visual assessment and consensus review. \revised{The total time for segmentation annotation and refinement is about $>400$ hours, and the centerline annotation and refinement takes about $>200$ hours.}

\revised{\bfsection{Clinical feasibility evaluation}
One of the most important downstream tasks of rib segmentation and centerline extraction is to facilitate fracture diagnosis. Hence, to ensure the clinical feasibility of the annotations, for all cases containing fracutres, we ensure all fractures in the case are 1) covered by the rib segmentation, and 2) passed through by the centerlines.

Specifically, for cases containing rib fractures, we consider the rib segmentation to be clinically feasible if it covers at least 75\% of each rib fracture segmentation. The threshold is loosened to be 75\% since the fracture segmentation from \textit{RibFrac} Dataset contains the region surrounding the fracture region, which includes the non-rib regions. While the clinical feasibility of centerline annotations are based on radiologists' visual assessment to ensure each centerline line passes through the corresponding rib fracture segmentation.
}


\subsection{Challenging Cases}\label{sec:method_challenge}
We report specific challenges of rib labeling and centerline extraction by analyzing and categorizing the abnormal cases, whose modalities are relatively rare. In clinical, however, the diagnosis of these cases is time-consuming, while for normal cases, even computer-assisted intervention is less needed. Hence, the discussion and categorization of these cases are valuable. Based on our visual assessment, 99 cases in \emph{RibSeg v2} are categorized as challenging cases (47/19/33 in training/development/test), which is contained in the dataset description file.

\bfsection{Challenging case categories}
We categorized 4 challenging situations: 1) The adjacent bones are connected by the growing callus or metal implants, as depicted in Fig.~\ref{fig:challenging_cases}~(a). Algorithmically, such cases will also cause morphological false merge, \ie, a single connected component contains several ribs. 2) The cases with metal implants like Fig.~\ref{fig:challenging_cases}~(a) also tend to suffer severe HU deviation, i.e., the HU value of the bone is higher/lower than their normal HU value. In such cases, the rib cage is wrapped by a 'noise shell', which is hard to filter. 3) The cases that are partly scanned or suffer severe bone lesions such as scoliosis in Fig.~\ref{fig:challenging_cases}~(b) and fractures like Fig.~\ref{fig:challenging_cases}~(c), which are hard to segment and label. 4) The floating ribs are missing or too vague to segment, as depicted in Fig.~\ref{fig:challenging_cases}~(d), and there might also exist a third stubby little floating rib (the 13th pair of ribs).



\section{Methodology}

\begin{figure*}[tb]
    \centering
	\includegraphics[width=\linewidth]{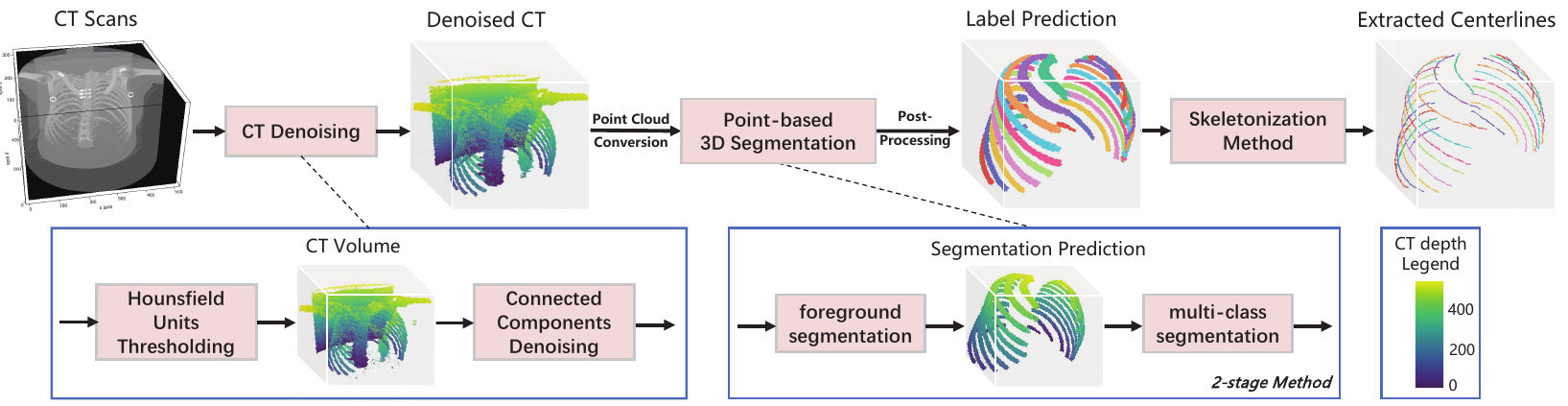}
	\caption{\textbf{The Pipeline of Rib Labeling and Anatomical Centerline Extraction.} The pipeline is divided into 3 steps: 
	1) \textbf{CT Denoising:} CT scans are thresholded by HU value and denoised by \emph{Connected-Component-Denoising} (CCD) to remove most non-bone voxels. 2) \textbf{Point-based Rib Labeling:} The denoised CT volume is converted to point clouds for a two-stage point-based segmentation method, which first predicts binary rib segmentation, and then segments individual ribs with a second model. The label prediction is further denoised by CCD for centerline extraction. 3) \textbf{Rib Centerline Extraction:} The rib centerlines are extracted from the label prediction via skeletonization methods. \textbf{Color:} For clear visualization, all binary volumes are colored by axial depth.} \label{fig:pipeline}
\end{figure*}



\subsection{Pipeline Overview}
\revised{Note that \emph{RibSeg v2} contains 2 tasks: 1) rib labeling and 2) anatomical centerline extraction from CT scans. Considering that rib centerlines can be easily obtained from rib labeling segmentation of high quality, we benchmark \emph{RibSeg v2} with a single pipeline that first segments individual ribs from CT scans and then extract centerlines from the segmentation. Specifically, rib labeling is formulated as a multi-class segmentation task and centerlines are extracted using skeletonization-based methods.}

As depicted in Fig.~\ref{fig:pipeline}, the pipeline is divided into three steps. 1) CT denoising: we first preprocessed the input CT scans to obtain the denoised CT volumes through morphological methods. 2) Point-based rib labeling: we converted the CT volume to point clouds and applied a two-stage point-based method to first obtain the binary rib segmentation and then segment individual ribs with a second model. Then the resultant label prediction is post-processed for centerline extraction. 3) Rib centerline extraction: based on the labeled segmentation, the centerlines are extracted through skeletonization methods.

\subsection{CT Denoising}
Considering the sparsity of ribs in 3D volumes ($< 0.5\%$ voxels) and the high HU value of bones in CT scans (${>200}$\,HU), we filter the non-target parts of CT volumes in a coarse-to-fine manner. Specifically, we first filter the non-bone voxels roughly by setting a threshold of 200 HU on CT volumes, which is the common CT attenuation value for bones. Although the resultant binarized volumes may contain many noises covering the rib cage, we keep the noises and propagate the volumes to the model training procedure to improve model robustness. While in the inference stage, we remove most noises by sorting out and eliminating the connected components of small volumes. We denote such connected components-based denoise procedure as \emph{Connected-Component-Denoising} (CCD). Note that CCD is crucial to obtaining high-quality rib labels, especially for the cases suffering HU deviation where the roughly filtered volumes will contain a huge number of noises. 

\subsection{Point Cloud Baseline for Rib Labeling}
\bfsection{Problem formulation}
We formulate it as a 25-class part segmentation problem to segment and label ribs from CT scans (24 classes for 24 ribs and 1 class for other bones and backgrounds). However, in this scenario, the target parts (24 ribs) are extremely sparse in the input volume ($< 0.7\%$ voxels after HU thresholding), which is different from conventional part segmentation tasks such as PASCAL-Part\cite{Chen2014DetectWY} and PartNet\cite{Mo2019PartNetAL}, where the segmentation parts of target objects have a rather balanced distribution. 

\bfsection{Frameworks}
To alleviate the sparsity issue, we propose a two-stage framework for rib labeling: 1) first perform binary segmentation to obtain ribs from CT scans, and 2) perform multi-class segmentation to segment individual ribs from binary segmentation. For comparison, we also test the one-stage method which directly predicts rib labels from CT volumes via multi-class segmentation in Sec.~\ref{subsec:exp_lab}.



\bfsection{Data representation for CT scans}
Most learning-based methods model CT scans as 3D volumes, and work on dense voxel grids, which is computationally inefficient\cite{lenga2018deep,wu2012learning}. To address the memory issue, we convert the dense 3D volumes to sparse point clouds\cite{yang2021ribseg} and adopt point-based networks as the backbone model. Specifically, we tested our benchmark pipeline with PointCNN\cite{Li2018PointCNNCO}, DGCNN\cite{Wang2018DynamicGC}, different input settings of PointNet\cite{Qi2016PointNetDL} and PointNet++\cite{qi2017pointnet++}. Note that PointCNN and DGCNN require a relatively high memory usage and could only afford 2048 points input at most, while PointNet and PointNet++, we also tested 30k points as input. For comparison, we also tested the voxel-based method with nnU-Net\cite{Isensee2020nnUNetAS} in Sec.~\ref{subsec:exp_lab}.

\begin{figure}[t]
    \centering
	\includegraphics[width=\linewidth]{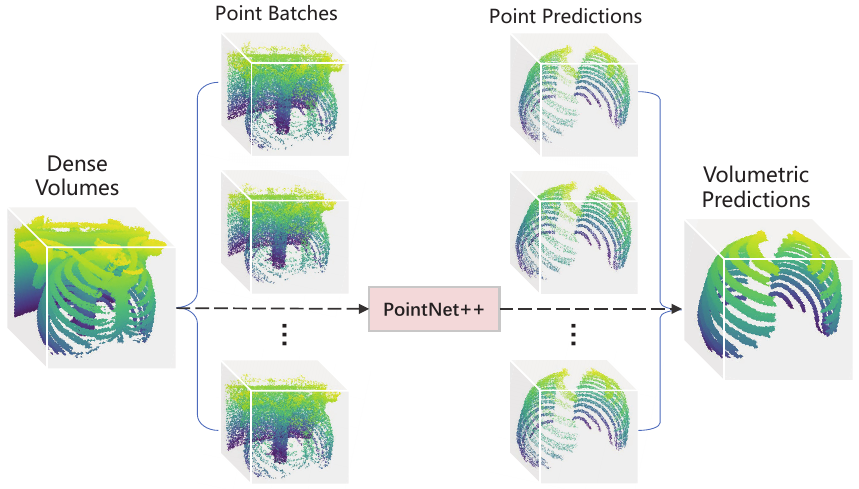}
	\caption{\textbf{Point clouds generation from dense volumes}. The CT volumes are converted to a point cloud and divided into equally-sized point batches as input, and the multi-batch point predictions are concatenated into volumetric predictions.} \label{fig:rep_pc}
\end{figure}

\bfsection{Point clouds conversion} In Fig.~\ref{fig:rep_pc}, we take binary rib segmentation as an example to show the procedure of point cloud conversion. Specifically, we first convert the CT volumes into a dense point cloud and divide the point cloud into equally-sized batches of points. For the batch with insufficient points, we 'ceil' it up by randomly sampling points from other batches, and applying majority voting to the prediction of repeated points. Finally, the point predictions of all batches are concatenated to obtain the voxel prediction.

\subsection{Skeletonization Methods for Rib Anatomical Centerline Extraction}
\revised{
During the annotation procedure, we found that both TEASAR-based method and L1-medial skeletonization method can guarantee visually satisfying results as long as the segmentation is correctly labeled. Hence, we simply cascade the skeletonization methods to the learning-based pipeline as an end-to-end baseline method for centerline extraction.
}

\bfsection{Voxel-based TEASAR method}
TEASAR method directly works on volumes. Specifically, 1) we first apply morphological operations to eliminate the mislabeled regions, and obtain connected components of individual ribs according to the volume. 2) Then for each rib, a raster scan is applied to locate an arbitrary foreground voxel, and its furthest point is denoted as the root point (it lies on the end of the connected component). 3) By implementing the Euclidean distance transform, a penalty field is defined\cite{Bitter2001penalizedis} to guide the centerline passing through the center of the rib volume. 4) Then Dijkstra's shortest path is implemented to derive the path from the root point to the most geodesically distant point from it (it lies on the other end of the connected component), and the resultant path is the extracted rib centerline. 5) Finally, we smoothen the result and upsample the centerline to 500 points by linear interpolation for the convenience of evaluation. 

\revised{
\bfsection{Point-based L1-medial skeletonization method}
L1-medial skeletonization method works on the point clouds. Specifically, in our case, we first 1) convert the dense volume into sparse point clouds. 2) Then we adopted the method in \cite{Huang2013L1medialSO} that adds a regularization term to L1-median to prevent the formation of point clusters and uses classical weighted PCA for skeleton branch detection. 3) And the curve skeleton can be generated by iterative contraction where the L1-medians of the local neighborhoods are represented by the point sets. 4) The resultant curves are smoothened and upsampled to 500 points by linear interpolation.

\bfsection{Shortest path-based method}
We also include a trivial shortest path-based method for comparison. Specifically, for each rib volume, we first select its endpoints according to coordinates on the transverse plane and perform the shortest path method used in \cite{Teng2011AutomatedPO}. The resultant point sets are also smoothened and upsampled to 500 points by linear interpolation.
}

\revised{
\subsection{Data Eligibility}\label{sec:eligibility}
The 654 of 660 CT volumes from \textit{RibFrac} dataset are included in this study, and the 6 unqualified cases are partly scanned and only cover the middle part of the ribcage, which makes it impossible to label the ribs. All 654 cases are used in the method development.

Note that all CT scans might have different level of HU deviation due to the hardware settings, so the 200 HU thresholding might filter foreground voxels and make the resultant ribs hollow inside. During the dataset development, we added morphological closing to ensure the completeness of the annotation. While for segmentation method, the converted point clouds still preseverve the sufficient geometry of ribs, which indicates that the HU thresholding won't limit the clinical applicability of the pipeline. Hence, all 654 cases are included in the pipeline development.
}

\revised{\subsection{Method Discussion}\label{sec:method_diss}
Clinically, rib anatomical centerlines are more favorable for being used to construct inner coordinate systems for surgery planning, while rib cage labeling is often utilized by visualization tools for structure and morphology analysis. In this study, we implement both two tasks in one pipeline where centerlines are extracted by applying skeletonization methods to the labeled ribs. Although the skeletonization-based method can achieve high-quality centerlines for most cases, the time consumption is not cheap (over 80s for a case), and it's sensitive to the quality of rib label predictions, which urges a more computationally efficient and robust method. 
}

\section{Experiments}\label{sec:experiments}
\subsection{Experiments on Rib labeling}\label{subsec:exp_lab}

\begin{figure*}[h]
    \centering
	\includegraphics[width=\linewidth]{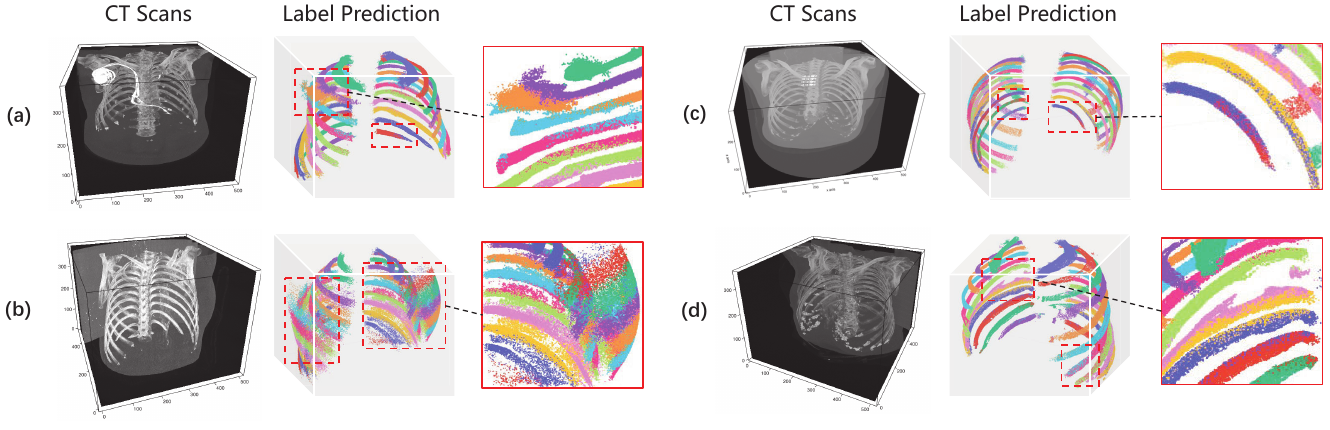}
	\caption{\textbf{Visualization of Label Predictions on Challenging Cases}. (a) The case contains a pacemaker, which partly remains in the label prediction. (b) The case suffers HU deviation, leading to severe cross-labeling. (c) The case misses the 12th pair of floating ribs, causing the cross-labeling. (d) The case has a serious fracture in the left 5th rib, causing the cross-labeling.} \label{fig:label_vis}
\end{figure*}

\subsubsection{Evaluation Metrics}
We first define the $\operatorname{Label-Dice}$\cite{Dice1945MeasuresOT, Taha2015MetricsFE} of rib $i$ as:
\begin{equation}
\operatorname{Dice}^{(L)}_{i}=  \frac{2\cdot |y_i\cdot \hat{y}_i|}{|y_i|+|\hat{y}_i|},
\end{equation} 
where $y$ and $\hat{y}$ indicate the label prediction and ground truth, respectively. For quantitative analysis, we evaluate the performance by reporting the average $\operatorname{Label-Dice}$ of the 24 ribs, denoted as $\operatorname{Dice}^{(L)}_{\text{avg}}$. While in qualitative analysis, we reflect the performance degradation by reporting the minimal $\operatorname{Label-Dice}$ amongst the 24 ribs, denoted as $\operatorname{Dice}^{(L)}_{\text{min}}$. Moreover, we report the $\operatorname{Label-Accuracy}$ of individual ribs to evaluate the method's clinical applicability. Specifically, an individual rib $i$ is counted as correctly labeled if \revised{$\operatorname{Recall}_{i}>0.7$}, and the accuracy can be calculated with ease. Considering that labeling first and twelfth rib pairs tend to be more difficult as they are shorter and curvier than other ribs, we report the $\operatorname{Label-Accuracy}$ of all / first / intermediate / twelfth rib pairs, respectively.



\subsubsection{Quantitative Analysis}
We first evaluate the methods on \emph{RibSeg v2} test set, comparing the one-stage and two-stage methods with different settings. As depicted in Tab.~\ref{tab:label_metrics}, two-stage methods significantly outperform one-stage methods \revised{with 0.2\%~$\sim$~16.5\% higher Label Dice (9.9\% on average) and 2.7\%~$\sim$~33.6\% higher label accuracy (13.7\% on average)}. The interpretation is that the rib in the input volume of the one-stage method is too sparse ($\sim$16\%) to provide sufficient features while the two-stage method removes most background noises for the label prediction. \revised{Due to the memory requirement, PointCNN and DGCNN could only afford 2048 points input, while for the light models (PointNet and PointNet++), we tested both 2048 and 30k points input for comparison, and the result indicated that increasing input size wouldn't necessarily improve the performance.} Besides, the point-based methods outperform the voxel-based method (\revised{nnU-Net}), as it inputs the geometry of the whole volume instead of voxel patches, which leads to a rich feature representation. \revised{Specifically, the best point-based method (DGCNN) enjoys 6.9\% higher label Dice values and 4.0\% higher label accuracy than nnU-Net. }

For inference speed, the point-based method is significantly more efficient for taking sparse point clouds as inputs, as reported in Tab.~\ref{tab:label_speed}, whereas the point-based method is \revised{${3\sim70\times}$} faster than the voxel-based method.



\begin{table}[h]
\centering
\caption{\textbf{Rib Labeling Metrics on \emph{RibSeg v2} Test Set}. \revised{The metrics include average $\operatorname{Label-Dice}$ and $\operatorname{Label-Accuracy}$ of all / first / intermediate / twelfth rib pairs (A/F/I/T). For model comparision, we adopted nnU-Net as the SOTA voxel-based model, and point-based models such as PointCNN, DGCNN, PointNet and PointNet++. Due to the memory requirement of PointCNN and DGCNN, all models take 2048 points as input. For the light PointNet and PointNet++, we further use 30k points input for comparision. Both one-stage and two-stage methods are included.}}
\label{tab:label_metrics}
\begin{tabular}{@{}ll|c|c@{}}
\toprule
\multicolumn{2}{c|}{Methods}        & $\operatorname{Dice}^{(L)}_{\text{avg}}$ & $\operatorname{Label-Accuracy}$ (A/F/I/T) \\ \midrule
\multicolumn{2}{c|}{\revised{nnU-Net}}        & 83.6\% & 87.5\% / 92.9\% / 87.7\% / 78.5\% \\  \midrule
\multicolumn{1}{c|}{\multirow{4}{*}{\rotatebox{90}{One-stage}}} & \revised{PointCNN} & 67.3\% & 55.5\% / 90.1\% / 51.5\% / 61.9\%  \\ 
\multicolumn{1}{l|}{} & \revised{DGCNN}      & 76.4\% & 77.4\% / 93.6\% / 76.8\% / 64.9\% \\ 
\multicolumn{1}{l|}{} & \revised{PointNet}   & 70.3\% & 66.7\% / 79.2\% / 66.0\% / 60.0\% \\ 
\multicolumn{1}{l|}{} & \revised{PointNet (30k)}   & 71.5\% & 70.1\% / 84.0\% / 68.9\% / 66.8\% \\ 
\multicolumn{1}{l|}{} & PointNet++ & 72.0\% & 70.0\% / 89.4\% / 68.6\% / 62.6\% \\
\multicolumn{1}{l|}{} & PointNet++ (30k) & 73.4\% & 72.0\% / 89.4\% / 70.9\% / 64.9\% \\ \midrule
\multicolumn{1}{r|}{\multirow{4}{*}{\rotatebox{90}{Two-stage}}} & \revised{PointCNN} & 83.8\% & 89.1\% / 54.3\% / 92.8\% / 87.2\% \\  
\multicolumn{1}{l|}{} & \revised{\textbf{DGCNN}}      & \textbf{90.5\%} & \textbf{91.5\%} / \textbf{60.7\%} / \textbf{96.3\%} / \textbf{89.4\%} \\  
\multicolumn{1}{l|}{} & \revised{PointNet}   & 75.1\% & 73.7\% / 50.6\% / 75.9\% / 74.3\% \\
\multicolumn{1}{l|}{} & \revised{PointNet (30k)}   & 71.7\% & 67.1\% / 47.0\% / 69.2\% / 67.2\% \\ 
\multicolumn{1}{l|}{} & PointNet++ & 84.4\% & 85.4\% / 59.1\% / 87.9\% / 87.9\% \\
\multicolumn{1}{l|}{} & PointNet++ (30k) & 84.8\% & 85.9\% / 59.4\% / 88.4\% / 87.5\% \\ \bottomrule
\end{tabular}
\end{table}

\begin{table}[tb]
\centering
\caption{\textbf{Speed Comparison}. The table reports model forward time in seconds (one-stage / two-stage) averaged from 10 sample cases. Post-processing time is not included as it heavily depends on the implementation.}
\label{tab:label_speed}
\begin{tabular}{@{}l|l@{}}
\toprule
Methods         & Forward Time (s) \\ \midrule
\revised{nnU-Net}         &    72.60 $\pm$ 17.19 / -               \\ \midrule
\revised{PointCNN}        &    20.21 $\pm$ 5.34 / 24.68 $\pm$ 5.67              \\
\revised{DGCNN}           &    11.24 $\pm$ 3.90 / 13.82 $\pm$ 4.11              \\
\revised{PointNet}        &    ~0.34  $\pm$ 0.40 / ~0.53 $\pm$ 0.55              \\
\revised{PointNet (30k)}   &    ~0.71  $\pm$ 1.82 / ~0.52 $\pm$ 0.58              \\
PointNet++      &    ~7.66  $\pm$ 1.90 / ~9.25 $\pm$ 1.95              \\
PointNet++ (30k) &    ~0.98  $\pm$ 0.50 / ~1.40 $\pm$ 0.64              \\ \bottomrule
\end{tabular}
\end{table}

\subsubsection{Qualitative Analysis}
For robustness analysis, we compare the inference results of normal and challenging cases and visualize the predictions on challenging cases.

\bfsection{Analysis on challenging cases} To evaluate the robustness of the method, we tested the best model on all/normal/challenging cases from the test set, respectively, as reported in Tab.~\ref{tab:label_metrics_anc}. The model enjoys a state-of-the-art performance in the normal cases while suffering a significant drop in the challenging cases: \revised{14.4\%} lower on average $\operatorname{Label-Dice}$ and \revised{14.4\%} lower on $\operatorname{Label-Accuracy}$. To further investigate the performance degradation, we also analyze the minimal instance-wise $\operatorname{Label-Dice}$, which is unsatisfying even in normal cases (\revised{73.4\%}) and suffers a huge drop by \revised{11.0\%} in the challenging cases. Note that the minimal $\operatorname{Label-Dice}$ occurs on rib 12 or 24 (the 12th pair of ribs). It is exactly these challenging cases that are clinically time-consuming to diagnose, and the floating ribs with various lesions are algorithmically difficult to segment, hence, a more robust method of tackling the challenging cases is desired.

\bfsection{Visualization of performance degradation} As the metrics may not necessarily reflect the prediction quality in detail, we further visualize the results in challenging cases in Fig.~\ref{fig:label_vis} (PointNet++). Specifically, for cases where adjacent ribs are connected by metal implants such as Fig.~\ref{fig:label_vis}~(a), where left 2 and 3 ribs are connected to a pacemaker, the prediction suffers serious cross-labeling. For cases that suffer HU deviation like Fig.~\ref{fig:label_vis}~(b), the prediction suffers cross-labeling and contains too many noises. For cases missing the floating ribs as Fig.~\ref{fig:label_vis}~(c), the model seems to impose 24 classes of rib labels on the rib cage, and the false ribs (the 8th to 12th pairs of ribs) are mislabeled. While in Fig.~\ref{fig:label_vis}~(d), the rib 5 left is severely damaged, also causing cross-labeling. In brief, despite the visually satisfying predictions in most cases, the performance degradation in a few challenging cases is significant. Considering the clinical practicality, it urges a more robust method for rib labeling with the ground truth provided by the \emph{RibSeg v2} dataset.

\begin{table}[tb]
	\caption{\textbf{Performance Comparison on Challenging Cases}. The baseline method is tested on all / normal / challenging cases from \emph{RibSeg v2} test set, respectively. The metrics include average $\operatorname{Label-Dice}$, minimal $\operatorname{Label-Dice}$, and $\operatorname{Label-Accuracy}$ over all pairs of ribs.} \label{tab:label_metrics_anc}
	\centering
	\begin{tabular*}{\hsize}{@{}@{\extracolsep{\fill}}l|c|c|c @{}}
		\toprule   
		Cases & $\operatorname{Dice}^{(L)}_{\text{avg}}$ & $\operatorname{Dice}^{(L)}_{\text{min}}$ & $\operatorname{Label-Accuracy}$ \\
		\midrule
		All & 90.5\% & 72.3\% & 91.5\% \\
		Normal & 92.6\% & 73.4\% & 94.0\% \\
		Challenging & 78.2\% & 62.4\% & 79.6\% \\
		\bottomrule
	\end{tabular*}
\end{table}

\subsection{Experiments on Rib Centerline Extraction}

\textbf{\begin{figure*}[tb]
    \centering
	\includegraphics[width=\linewidth]{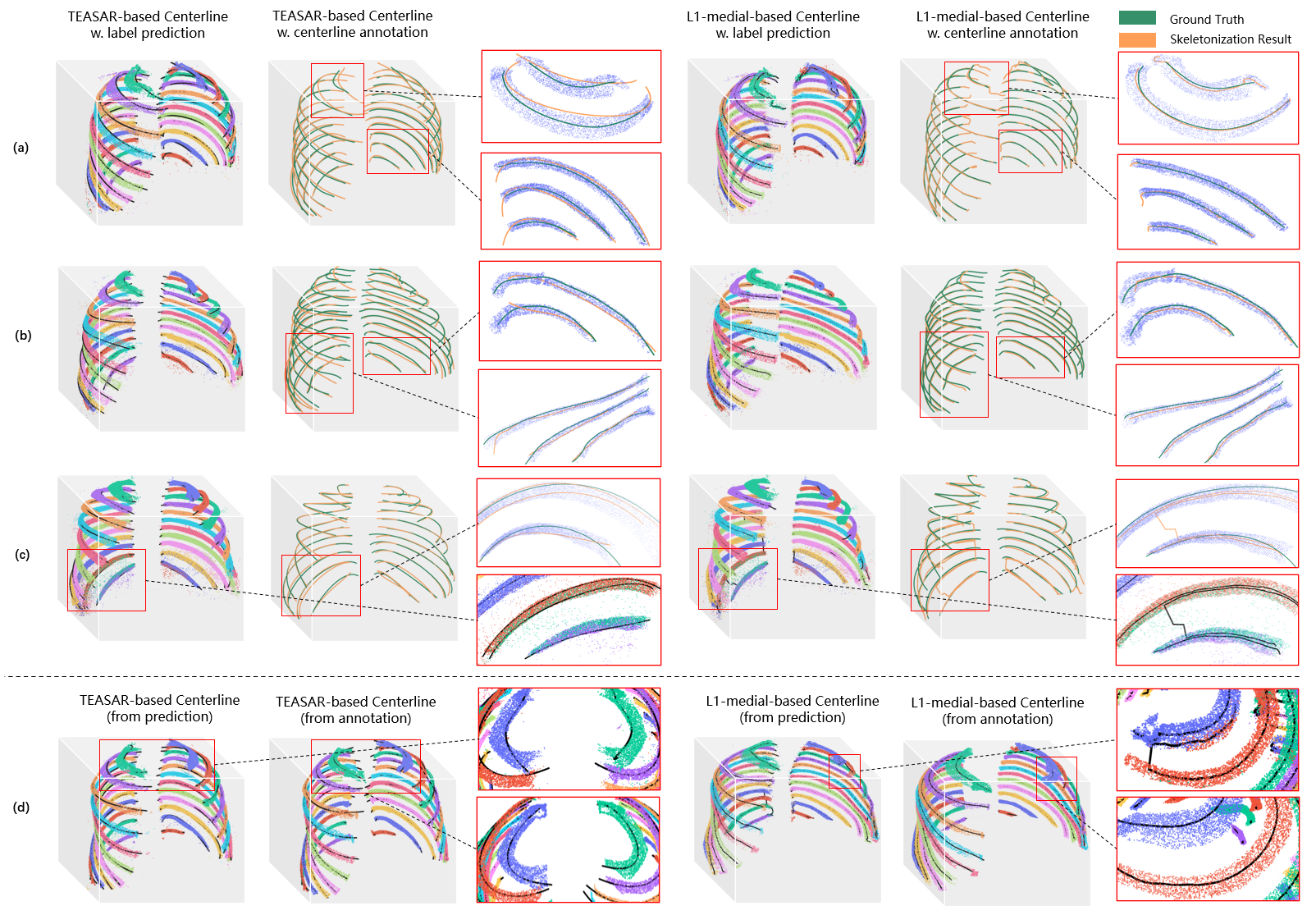}
	\caption{\textbf{Visualization of Centerline Extraction Results}. \revised{(a$\sim$c) Evaluation on rib centerline extraction pipeline: TEASAR method and L1-medial skeletonization are applied to rib label predictions and the resultant centerlines are compared with the centerline annotation. Shortest path method fails on most cases, which is not included. (d) Ablation study on skeletonization methods: both methods are applied to rib label predictions and annotation, respectively.}} \label{fig:cl_vis}
\end{figure*}}

\label{subsec:exp_cl}
\subsubsection{Evaluation Metrics} \revised{For centerline evaluation, we used three metrics to measure 1) the similarity/distance between the extracted centerline and annotation, 2) the curvature deviation and 3) the axial deviation of the extracted centerline. We first adopted $\operatorname{centerlineDice}$ ($\operatorname{clDice}$)\cite{Shit2020clDiceA} to evaluate the rib-wise distance between extracted centerline and the centerline annotation, and denote the $\operatorname{clDice}$ of rib $i$ as:
\begin{equation}
\operatorname{clDice}_{i}=2 \times \frac{T_{\text{prec}}\left(L_{i}, \hat{S}_{i}\right) \times T_{\text{sens}}\left(\hat{L}_{i}, S_{i}\right)}{T_{\text{prec}}\left(L_{i}, \hat{S}_{i}\right)+T_{\text{sens}}\left(\hat{L}_{i}, S_{i}\right)},
\end{equation}
where $L_{i}$, $\hat{L}_{i}$, $S_{i}$, and $\hat{S}_{i}$ indicate the centerline prediction, centerline annotation, label prediction, and label annotation of the rib $i$, respectively. While $T_{\text{prec}}$ and $T_{\text{sens}}$ indicate the topology’s precision and sensitivity, which are defined as:
\begin{equation}
T_{\text{prec}}\left(L_{i}, \hat{S}_{i}\right)=\frac{\left|L_{i} \cap \hat{S}_{i}\right|}{\left|L_{i}\right|} ; \quad T_{\text{sens}}\left(\hat{L}_{i}, S_{i}\right)=\frac{\left|\hat{L}_{i} \cap S_{i}\right|}{\left|\hat{L}_{i}\right|}.
\end{equation}
For quantitative analysis, we report the average $\operatorname{clDice}$ of the 24 ribs, denoted as $\operatorname{clDice}_{\text{avg}}$.}

Inspired by normalized surface dice~\cite{Niko2018nsd,Zou2004dsc}, we propose \emph{Normalized-Line-Dice} ($\operatorname{NLD}$) to evaluate the curvature deviation:
\begin{equation}
\centering
\operatorname{NLD}(\mathbf{L}, \mathbf{\hat{L}}) = \frac{| \mathbf{L} \cap B_{\mathbf{\hat{L}}}^{(\tau)}|+|  \mathbf{\hat{L}}
\cap B_{ \mathbf{L}}^{(\tau)}|}{| \mathbf{L}|+|\mathbf{\hat{L}}|},
\end{equation}
where $\mathbf{L} \subseteq \mathbb{R}^{3}$ and $\mathbf{\hat{L}} \subseteq \mathbb{R}^{3}$ are the extracted centerline and centerline annotation, respectively, and $B_{\mathbf{L}}^{(\tau)}=\{y \in \mathbb{R}^{3} \mid \exists ~ \tilde{y} \in  \mathbf{L},\|y-\tilde{y}\|_{2} \leq \tau\}$ denotes the surrounding region of the centerline $\mathbf{L}$ within tolerance distance $\tau$. Here we take $\tau=7$, which is roughly the radius length of the cross surface for an approximate rib cylinder.


\revised{Another important measurement is axial deviation, indicating whether the centerline lies in the mid of the corresponding rib bone. Hence, we report the unidirectional Chamfer Distance of the extracted centerline with respect to the annotation of rib segmentation, denoted as \emph{Line-Seg-Chamfer-Distance} ($\operatorname{LSCD}$)}:
\begin{equation}
     \operatorname{LSCD}(\mathbf{L}, \mathbf{\hat{S}})=
     \frac{1}{|\mathbf{\hat{S}}|} \sum_{\hat{y} \in \mathbf{\hat{S}}} \min _{y \in \mathbf{L}}\|y-\hat{y}\|_{2},
\end{equation}\label{eq:lscd}
where $\mathbf{L} \subseteq \mathbb{R}^{3}$ and $\mathbf{\hat{S}} \subseteq \mathbb{R}^{3}$ are the extracted centerline and the annotation of rib labels, respectively. 
\revised{
Since Chamfer distance is not numerically intuitive, we report the relative error between $\operatorname{LSCD}$ of extracted centerline and $\operatorname{LSCD}$ of centerline annotation, denoted as \emph{Line-Seg-Chamfer-Distance-Error} ($\operatorname{LSCDError}$):

\begin{equation}
     \operatorname{LSCDError}(\mathbf{L}, \mathbf{\hat{L}})=
     \frac{\operatorname{LSCD}(\mathbf{L}, \mathbf{\hat{S}}) - \operatorname{LSCD}(\mathbf{\hat{L}}, \mathbf{\hat{S}})}{\operatorname{LSCD}(\mathbf{\hat{L}}, \mathbf{\hat{S}})}.
\end{equation}\label{eq:lscde}

}



\subsubsection{Quantitative Analysis} 
We first applied the skeletonization methods to the rib label prediction with the highest accuracy (\revised{two-stage DGCNN}) from Sec.~\ref{subsec:exp_lab}. As reported in Tab.~\ref{tab:cl_metrics}, we evaluated TEASAR method, \revised{L1-medial skeletonization} and \revised{shortest path method} based on \revised{$\operatorname{clDice}$}, $\operatorname{NLD}$, and \revised{$\operatorname{LSCDError}$}, respectively. \revised{Since the skeletonization methods might be sensitive to the rib label predictions from the first stage, we also directly applied these methods to the rib label annotations as an ablation study.

Overall, given the rib label prediction with noises, L1-medial skeletonization performs the best, with a 8.0\% higher $\operatorname{clDice}$, a close $\operatorname{NLD}$ (0.7\% lower), and a 24.3\% lower $\operatorname{LSCDError}$ than TEASAR method. When applied to the rib label annotations, both L1-medial skeletonization and TEASAR method obtained results with very high accuracy, where L1-medial skeletonization performs slightly better than TEASAR method (2.4\% higher $\operatorname{clDice}$, 1.3\% higher $\operatorname{NLD}$, and 0.8\% $\operatorname{LSCDError}$). And we concluded that such non-learning skeletonization methods can work well on rib centerline extraction task, especially when applied to high quality rib segmentation. 

However, the trivial shortest path fails to give a good result even given the rib label annotation, where the resultant path always deviates from the actual centerline. When applied to rib label predictions, it fails on the majority cases and we didn't include its results in Tab.~\ref{tab:cl_metrics}. The interpretation is that it requires the rib volumes to be one single connected component while the rib label predictions contain too many noises, and the discrete noise points could be detected as ending points, which fails the algorithm.}

\begin{table}[h]
\centering
\caption{\textbf{Rib Centerline Extraction Metrics on \emph{RibSeg v2} test set}. We include the ablation study of applying skeletonization to rib label predictions (p) and rib annotations (a), respectively (p / a). The methods include l1-medial skeletonization, TEASAR method, and a trivial shortest path. The metrics include \emph{centerlineDice} ($\operatorname{clDice}$), \emph{Normalized-Line-Dice} ($\operatorname{NLD}$), and \emph{Line-Seg-Chamfer-Distance-Error ($\operatorname{LSCDError})$}.}
\label{tab:cl_metrics}
\begin{tabular}{@{}l|l|l|l@{}}
\toprule
Methods       & \revised{$\operatorname{clDice}$}          & $\operatorname{NLD}$             & \revised{$\operatorname{LSCDError}$}           \\ \midrule
\revised{L1-medial}     & \textbf{90.8\%} / \textbf{97.4\%} & 82.1\% / \textbf{97.9\%} & \textbf{38.8\%} / \textbf{8.4\%} \\ \midrule
TEASAR        & 82.8\% / 95.0\% & \textbf{82.8\%} / 96.6\% & 63.1\% / 9.2\% \\ \midrule
\revised{Shortest path} & - / 33.5\% & - / 55.9\% & - / 20.0\% \\ \bottomrule
\end{tabular}
\end{table}

\subsubsection{Qualitative Analysis}
\revised{As depicted in Fig.~\ref{fig:cl_vis}, we visualized the centerline extraction results for a more intuitive and detailed analysis. In most cases, with accurate label predictions, both skeletonization methods can guarantee visually satisfying centerlines. As depicted in Fig.~\ref{fig:cl_vis}~(a), the resultant centerline of the voxel-based TEASAR method exceeds the actual rib a little bit and slightly deviates from the annotation, the interpretation is that the TEASAR method includes dilation which dilates the volume making the extracted centerline longer, and also enlarges the noise voxels hence dislocating the centroid. While the point-based L1-medial skeletonization is more shape-sensitive and the resultant centerline aligns the annotation better. Fig.~\ref{fig:cl_vis}~(b) shows a case containing obvious rib fracture, where both methods managed to generate a complete centerline passing through the fracture region. However, the skeletonization methods might suffer a huge performance drop if the prediction contains too much mislabeling. As depicted in Fig.~\ref{fig:cl_vis}~(c), this case misses a floating rib and the label prediction suffers a huge accuracy drop which heavily affected the skeletonization methods. Consequentially, the voxel-based TEASAR result contains 2 centerlines aligning to the same rib, while the point-based L1-medial skeletonization result contains a misaligned centerline crossing 2 ribs. Similarly, the ablation study in Fig.~\ref{fig:cl_vis}~(d) further indicates that the quality of centerline extraction heavily depends on the label prediction. For the voxel-based TEASAR method, the resultant centerline deviates from the rib center. While the point-based L1-medial is much more sensitive to the rib label prediction, as the cross-labeling regions in the prediction will misguide the resultant centerline. Moreover, as mentioned in Sec.~\ref{sec:method_challenge}, skeletonization methods also fail for cases suffering severe rib fracture, where a single rib is broken into several parts, which will result in messy curve segments. Such abnormal cases also urge a more robust centerline extraction method, with centerline annotations provided by \emph{RibSeg v2}.}


\subsubsection{Discussion on annotation and metrics}
During the experiment, we noticed that even though the centerline generated by our method perfectly lies in the center of the rib segmentation by visual assessment, it might not necessarily have the minimal $\operatorname{LSCDError}$ value. The interpretation is that annotations of rib anatomical centerlines are not perfect because manual annotation of the centerline is a naturally hard task for humans, and even for well-trained radiologists, it's difficult to locate endpoints that exactly lie in the center of the ribs before connecting them as the centerline curve. However, although being geometrically imperfect, such manually confirmed centerline annotations are still clinically pragmatic and valuable.

\subsection{Experiment Settings}
\revised{The training of the models was carried out on a cluster with 4 NVIDIA A100.} The inference was conducted with the implementation of PyTorch 1.7.1 and Python 3.9, on a machine with a single NVIDIA Tesla P100 GPU with Intel(R) Xeon(R) CPU @ 2.20 GHz and 150 G memory. In the training stage of the two-stage method, the input of the segmentation task is point sets of \revised{2,048} / 30,000 downsampled from the binary CT volume, while the input of the labeling task is point sets downsampled from the predictions of segmentation. For point-based models, the Adam optimizer is adopted and a combination of cross-entropy (CE) and Dice loss as the loss function. \revised{All experiments settings and trained models are available online at \hyperlink{https://github.com/M3DV/RibSeg}{https://github.com/M3DV/RibSeg}.}


\section{Conclusion}
We developed the \emph{RibSeg v2} dataset, which is the first open dataset for rib labeling and anatomical centerline extraction. Besides, we explored the challenges of rib labeling and centerline extraction in detail and benchmark \emph{RibSeg v2} with a strong pipeline including a deep learning-based method for rib labeling and \revised{skeletonization}-based methods for centerline extraction. We then compared data representations of CT scans as dense voxel grids and sparse point clouds and provided a comprehensive analysis of the abnormal cases where the method might fail. Besides, \revised{to evaluate our pipeline, we tested 4 point-based models with different settings and 3 skeletonization methods,} and also proposed various metrics for rib segmentation, labeling, and anatomical centerline extraction, providing a comprehensive method evaluation. Finally, by detailed quantitative and qualitative analysis of the categorized challenging cases, we featured the key challenges of each task, which are valuable to guide future studies.

The dataset and proposed method show the potential to be clinically applicable, such as the diagnosis of rib fractures and bone lesions. Besides, considering the differences from standard medical image datasets~\cite{antonelli2021medical,yang2020medmnist} with pixel/voxel grids, the elongated shapes and oblique poses of ribs enable the \emph{RibSeg v2} dataset to serve as a benchmark for curvilinear structures and geometric deep learning.

There remain limitations in this study. For rib anatomical centerline extraction, we apply a \revised{skeletonization}-based method to extract centerlines from the prediction of rib labels, which is sensitive to rib labeling errors in abnormal cases. Hence, considering the clinical significance of rib anatomical centerline extraction, a more robust method will be favorable.


\bibliographystyle{IEEEtran}
\bibliography{reference}

\end{document}